%% file: mydocument.tex
\newcommand*{\ATLASLATEXPATH}{latex/}
\documentclass[cernpreprint,texlive=2011,txfonts,UKenglish]{latex/atlasdoc} 

\usepackage{\ATLASLATEXPATH atlaspackage}
\usepackage{latex/atlaspackage}
\usepackage{\ATLASLATEXPATH atlasbiblatex}

\usepackage{\ATLASLATEXPATH atlascontribute}

\usepackage{\ATLASLATEXPATH atlasphysics}
\usepackage{latex/atlasphysics} 

\graphicspath{{logos/}{figures/}}

\usepackage{mydocument-defs}

\input{mydocument-metadata}
\hypersetup{pdftitle={ATLAS document},pdfauthor={The ATLAS Collaboration}}

\begin{document}

\maketitle



\section{Introduction}
\label{sec:intro}

The production of \wz pairs in hadron collisions is an important test of the electroweak sector of the Standard Model (SM). 
The \wz final states arise from two vector bosons radiated by quarks or from the decay of a virtual $W$ boson into a 
\wz pair, which involves a triple gauge coupling (TGC). In addition, \wz pairs can be produced 
in vector-boson scattering processes, which involve triple and quartic gauge couplings (QGC) and are 
sensitive to the electroweak symmetry breaking sector of the SM. New physics could manifest in \wz final 
states as a modification of the TGC and QGC strength. Precise knowledge of the \wz production cross section is therefore necessary in the search 
for new physics. 

Measurements of the \wz production cross section in proton--antiproton collisions
at a centre-of-mass energy of $\sqrt{s} = 1.96$~\TeV{} were published by
the CDF and D0 Collaborations~\cite{CDF:2012WZ,D0:2012WZ} using integrated luminosities of $7.1$~fb$^{-1}$ and $8.6$~fb$^{-1}$, respectively.
At the Large Hadron Collider (LHC), measurements have been performed in proton--proton ($pp$) collisions by the ATLAS Collaboration~\cite{Aad:2012twa,ATLASWZ_8TeV} 
at $\sqrt{s}= 7$~\TeV{} and $8$~\TeV{} using integrated luminosities of $4.6$~fb$^{-1}$ and $20.3$~fb$^{-1}$, respectively.

This Letter presents measurements of the \wz production cross section in $pp$ collisions at a
centre-of-mass energy of $\sqrt{s}$ = $13$~\TeV{}. The data sample analysed was collected in $2015$ by the ATLAS 
experiment at the LHC, and corresponds to an integrated luminosity of {\mbox {$3.2$~fb$^{-1}$}}.
The $W$ and $Z$ bosons are reconstructed using their decay modes into electrons or muons.
The inclusive production cross section is measured in a fiducial phase space and extrapolated to the total phase space.
This Letter also reports the ratio of the cross sections at 13~\TeV{} and 8~\TeV~\cite{ATLASWZ_8TeV}, as well as
the ratio of the $W^+Z$/$W^-Z$ cross sections, which is sensitive to the parton distribution functions (PDF).   
Finally, the production cross section is also measured as a function of the jet multiplicity.
This distribution provides an important test of perturbative quantum chromodynamics (QCD) for diboson production processes.
The \wz diboson process is particularly well suited for this measurement, since the $WW$ final state has a 
very large background from top-quark production when associated jets are present, and the $ZZ$ final state has substantially fewer events. 
The reported measurements are compared with the SM cross-section predictions at the next-to-leading order (NLO) in QCD~\cite{Ohnemus:1991,Frixione:1992} 
and the total cross section is also compared to a very recent calculation at next-to-next-to-leading order (NNLO) in QCD~\cite{Grazzini:2016swo}.

\section{ATLAS detector}
\label{sec:detector}

The ATLAS detector~\cite{ATLAS_Detector} is a multi-purpose detector with a cylindrical 
geometry\footnote{ATLAS uses a right-handed coordinate system with its origin at the nominal 
interaction point (IP) in the centre of the detector and the $z$-axis along the beam direction. 
The $x$-axis points from the IP to the centre of the LHC ring, and the y-axis points upward. 
Cylindrical coordinates $(r,\phi)$ are used in the transverse $(x,y)$ plane, $\phi$ being the 
azimuthal angle around the beam direction. The pseudorapidity is defined in terms of the
polar angle $\theta$ as $\eta = -\textrm{ln}[ \textrm{tan}(\theta/2)]$. 
} and nearly $4\pi$ coverage in solid angle.
The collision point is surrounded by inner tracking detectors (collectively referred to as 
the inner detector), followed by a superconducting solenoid providing a $2$~T axial magnetic 
field, a calorimeter system and a muon spectrometer.

The inner detector (ID) provides precise measurements of charged-particle tracks in the pseudorapidity range $|\eta| < 2.5$.
It consists of three subdetectors arranged in a coaxial geometry around the beam axis: a silicon pixel detector, 
a silicon microstrip detector and a transition radiation tracker. 
The newly installed innermost layer of pixels sensors~\cite{IBL,ATLAS:1451888} was operational for the first time during the 2015 data taking.
  
The electromagnetic calorimeter covers the region $|\eta|<3.2$ and is based on a high-granularity, lead/liquid-argon (LAr) sampling technology. 
The hadronic calorimeter uses a steel/scintillator-tile detector in the region $|\eta|<1.7$ and a copper/LAr detector in the region $1.5 < |\eta| < 3.2$. 
The most forward region of the detector, $3.1 <|\eta| < 4.9$, is equipped with 
a forward calorimeter, measuring electromagnetic and hadronic energies in copper/LAr and tungsten/LAr modules.

The muon spectrometer (MS) comprises separate trigger 
and high-precision tracking chambers to measure the deflection of muons in a magnetic field generated by three 
large superconducting toroids arranged with an eightfold azimuthal coil symmetry around the calorimeters. 
The high-precision chambers cover a range of $|\eta|< 2.7$. 
The muon trigger system covers the range $|\eta|< 2.4$ with 
resistive-plate chambers in the barrel and thin-gap chambers in the endcap regions.

A two-level trigger system is used to select events in real time. It consists of a hardware-based first-level trigger and a software-based high-level trigger.
The latter employs algorithms similar to those used offline to identify electrons, muons, photons and jets.

\section{Phase space definition}
\label{sec:FiducialPS}

The fiducial phase space used to measure the \wz\ cross section is defined to closely follow the criteria 
used to define the signal region described in Section~\ref{sec:EventSelection}.
The phase space is based on the kinematics of the final-state leptons associated with the $W$ and $Z$ boson 
decays. Leptons produced in the decay of a hadron, a $\tau$ or their descendants are not considered in the 
definition of the fiducial phase space.
In the simulation, the kinematics of the charged lepton after quantum electrodynamics (QED) final-state radiation (FSR) are ``dressed'' 
at particle level by including contributions from photons with an angular distance $\Delta R \equiv \sqrt{(\Delta\eta)^2 + (\Delta\phi)^2} < 0.1$ 
from the lepton. Dressed leptons, and final-state neutrinos that do not originate from hadron or $\tau$ decays, 
are matched to the $W$ and $Z$ boson decay products using a Monte Carlo generator-independent algorithmic approach, 
called the ``resonant shape'' algorithm~\cite{ATLASWZ_8TeV}, that takes into account the nominal line shapes of the $W$ and $Z$ resonances.

The reported cross sections are measured in a fiducial phase space
defined at particle level by the following requirements:
the transverse momentum \pt\ of the dressed leptons from the $Z$ boson decay is 
above $15$~\GeV, the \pt of the charged lepton from the $W$ decay is above $20$~\GeV, 
the absolute value of the pseudorapidity of the charged leptons from the $W$ and $Z$ bosons is below $2.5$,
the invariant mass of the two leptons from the $Z$ boson decay differs at most by $10$~\GeV{} 
from the world average value of the $Z$ boson mass $m_Z^{\textrm{PDG}}$~\cite{Agashe:2014kda}. 
The $W$ transverse mass, defined as $m_{\textrm{T}}^{W} = \sqrt{2 \cdot p_{\rm{T}}^\nu \cdot p_{\rm{T}}^\ell \cdot [1 -\cos{\Delta \phi(\ell, \nu)}]}$, where $\Delta \phi(\ell, \nu)$ 
is the angle between the lepton and the neutrino in the transverse plane, is required to be above $30$~\GeV. 
In addition, it is required that the angular distance $\Delta R$ 
between the charged leptons from the $W$ and $Z$ decay is larger than $0.3$, and that
$\Delta R$ between the two leptons from the $Z$ decay is larger than $0.2$.

The fiducial cross section is extrapolated to the total phase space and corrected for the leptonic 
branching fractions of the $W$ and $Z$ bosons. The total phase space is defined by requiring the 
invariant mass of the lepton pair associated with the $Z$ boson to be in the range $66 < m_{\ell\ell} < 116$~\GeV.

For the jet multiplicity differential measurement, particle-level jets are reconstructed from stable particles 
with a lifetime of $\tau>30$~ps in the simulation after parton showering, hadronisation, and decay of particles with $\tau<30$~ps.
Muons, electrons, neutrinos and photons associated with $W$ and $Z$ decays are excluded. The particle-level jets are reconstructed with the anti-$k_{t}$ 
algorithm~\cite{antikt} with a radius parameter $R=0.4$ and are required to have a \pt above $25$~\GeV{} and an absolute value of pseudorapidity below $4.5$.

\section{Simulated event samples}
\label{sec:SimulatedSamples}

Monte Carlo (MC) simulation is used to model signal and background processes.
All generated MC events are passed through the ATLAS detector simulation~\cite{Aad:2010ah}, based on 
GEANT4~\cite{Agostinelli:2002hh}, and processed using the same reconstruction software used for the data.
The event samples include the simulation of additional proton--proton interactions (pile-up) 
generated with \PYTHIA~8.186~\cite{Sjostrand:2007gs} using the MSTW2008LO PDF~\cite{Martin:2009iq} 
and the A2~\cite{ATL-PHYS-PUB-2012-003} set of tuned parameters. 

Scale factors are applied to simulated events to correct for the small differences between data and MC simulation
in the trigger, reconstruction, identification, isolation and impact parameter efficiencies of electrons 
and muons~\cite{ATL-PHYS-PUB-2015-041,ATLAS-CONF-2014-032,Aad:2016jkr}. 
Furthermore, the electron energy and muon momentum in simulated events are smeared to account for small 
differences in resolution between data and MC~\cite{Aad:2014nim,Aad:2016jkr}.

A sample of simulated \wz\ events is used to correct the signal yield for detector effects, to extrapolate 
from the fiducial to the total phase space, and to compare the measurements to the theoretical predictions.
The production of \wz\ pairs and the subsequent leptonic decays of the vector bosons are generated at NLO in QCD using the 
\POWHEG-\textsc{Box}~v2~\cite{powheg1:2004, powheg, Alioli:2010xd, Melia:2011tj} generator, 
interfaced to the \PYTHIA~8.210 parton shower model using the AZNLO~\cite{AZNLO:2014} set of tuned parameters.
The CT10~\cite{CT10} PDF set is used for the hard-scattering process, 
while the CTEQ6L1~\cite{Pumplin:2002vw} PDF set is used for the parton shower.  
The jet multiplicity measurement is also compared to the theoretical NLO prediction from the \SHERPA 2.1.1 
generator~\cite{Sherpa}, calculated using the CT10 PDF set in conjunction with a dedicated set of tuned 
parameters for the parton shower developed by the \sherpa authors~\cite{Schumann:2007mg}.

The background sources in this analysis include processes with two or more electroweak gauge bosons, namely $ZZ$, 
$WW$ and $VVV$ ($V=W,Z$); processes with top quarks, such as $t\bar{t}$ and $t\bar{t}V$, single top and $tZ$; or processes with 
gauge bosons associated with jets or photons ($Z+j$ and $Z\gamma$).
MC simulation is used to estimate the contribution from 
background processes with three or more prompt leptons. Background processes with at least one misidentified 
lepton are evaluated using data-driven techniques and simulated events are used to assess the systematic uncertainties in these backgrounds.

The $q\bar{q} \to ZZ^{(\ast)}$, $t\bar{t}$, and single-top processes are generated at NLO using the 
\POWHEG-\textsc{Box}~v2 program. The CT10 PDF set is used for the matrix-element calculations. 
For the $ZZ$ process the parton shower is modelled with \PYTHIA~8.186, using the CTEQ6L1 PDF and AZNLO 
set of tuned parameters. The modelling of the parton shower for processes with top quarks is done with \PYTHIA~6.428~\cite{Sjostrand:2006za}, 
using the CTEQ6L1 PDF and Perugia 2012~\cite{Skands:2010ak} set of tuned parameters. 
The \SHERPA~\cite{Sherpa,Hoeche:2009rj,Gleisberg:2008fv,Schumann:2007mg,Schonherr:2008av,Hoche2013} event generator 
is used to model the $Z\gamma$, $VVV$, and $gg\to ZZ^{(\ast)}$ processes at leading order (LO) using the CT10 PDF set. 
Finally, the $t\bar{t}V$ and $tZ$ processes are generated at LO using \MGMCatNLO~\cite{Alwall:2014hca} with the 
NPDF23LO~\cite{Ball:2012cx} PDF set, interfaced with \PYTHIA~8.186 ($t\bar{t}V$) and \PYTHIA~6.428 ($tZ$). 

\section{Data sample and selections}
\label{sec:EventSelection}

The $pp$ collision data analysed correspond to an integrated luminosity of $3.2~\ifb$ collected with the ATLAS detector in 2015 at $\sqrt{s} = 13$~\TeV. 
Only data recorded with stable beam conditions and with all relevant detector subsystems operational are considered.

Candidate events are selected using triggers~\cite{ATL-DAQ-PUB-2016-001} that require at least one electron or muon with $\pt>24$~\GeV{} or 20~\GeV{}, respectively, 
that satisfies a loose isolation requirement. Possible inefficiencies for leptons with large transverse momenta are reduced by including additional electron and muon 
triggers that do not include any isolation requirements with transverse momentum thresholds of $\pT = 60$~\GeV{} and 50~\GeV{}, respectively. Finally, a single-electron 
trigger requiring $\pT>120$~\GeV{} with less restrictive electron identification criteria is used to increase the selection efficiency for high-$\pt$ electrons.

Events are required to have a primary vertex reconstructed from at least two charged particle tracks and compatible with the luminous region. 
If several such vertices are present in the event, the one with the highest sum of the $\pT^{2}$ of the associated tracks is selected as the primary vertex of the \wz\ production.

All final states with three charged leptons (electrons $e$ or muons $\mu$) and neutrinos from \wz\ leptonic decays are considered. In the following, the different final 
states are referred to as $\mu^{\pm}\mu^{+}\mu^{-}$, $e^{\pm}\mu^{+}\mu^{-}$, $\mu^{\pm}e^{+}e^{-}$ and $e^{\pm}e^{+}e^{-}$.

Muon candidates are identified by tracks reconstructed in the muon spectrometer and matched to tracks reconstructed in the inner detector. 
Muons are required to pass a ``medium'' identification selection, which is based on requirements on the number of hits in the ID and the 
MS~\cite{Aad:2016jkr}. The efficiency of this selection averaged over \pt and \eta is larger than $98\%$.  
The muon momentum is calculated by combining the MS measurement, corrected for the energy deposited in the calorimeters, and the ID measurement.
The $\pt$ of the muon must be greater than 15~\GeV{} and its pseudorapidity must satisfy $\abseta<2.5$. 

Electron candidates are reconstructed from energy clusters in the electromagnetic calorimeter matched to inner detector tracks. 
Electrons are identified using a discriminant that is the value of a likelihood function constructed with information from the 
shape of the electromagnetic showers in the calorimeter, track properties and track-to-cluster matching quantities of the candidate~\cite{ATL-PHYS-PUB-2015-041}. 
Electrons must satisfy a ``medium'' likelihood requirement, which provides an overall identification efficiency of 90\%.
The electron momentum is computed from the cluster energy and the direction of the track. The $\pt$ of the electron must be 
greater than 15~\GeV{} and the pseudorapidity of the cluster must be in the ranges $\abseta< 1.37$ or $1.52 < \abseta < 2.47$.

Electron and muon candidates are required to originate from the primary vertex. Thus, the significance of the track's transverse 
impact parameter calculated with respect to the beam line, $|{d_{0}/\sigma_{d_{0}}}|$, must be smaller than three for muons and 
less than five for electrons, and the longitudinal impact parameter, $z_{0}$ (the difference between the value of $z$ of the point 
on the track at which $d_{0}$ is defined and the longitudinal position of the primary vertex), is required to satisfy $|z_0\cdot \sin(\theta)|<0.5$~mm.

Electrons and muons are required to be isolated from other particles. The isolation requirement is based on both calorimeter and 
track information and is tuned for an efficiency of at least 95\% for $\pT>25$~\GeV\ and at least 99\% for $\pT>60$~\GeV~\cite{Aad:2016jkr}.

Jets are reconstructed from clusters of energy deposition in the calorimeter~\cite{PERF-2014-07} using the anti-$k_{t}$ algorithm~\cite{antikt} 
with a radius parameter $R=0.4$. Events with jets arising from detector noise or other non-collision sources are discarded~\cite{ATLAS-CONF-2015-029}. 
All jets must have $\pt>25$~\GeV{} and be reconstructed in the pseudorapidity range $|\eta|<4.5$. A multivariate combination of track-based 
variables is used to suppress jets originating from pile-up in the ID acceptance~\cite{Aad:2015ina}. The energy of jets is calibrated and corrected 
for detector effects using a combination of simulated events and {\textit{in situ}} methods in $13$~\TeV{} data, similar to the procedure described in Ref.~\cite{ATL-PHYS-PUB-2015-015}.

The transverse momentum of the neutrino is estimated from the missing transverse momentum in the event, \met, 
calculated as the negative vector sum of the transverse momentum of all identified hard physics objects 
(electrons, muons, jets), as well as an additional soft term. A track-based measurement of the soft term~\cite{ATL-PHYS-PUB-2015-027}, 
which accounts for low-$\pt$ tracks not assigned to a hard object, is used in the analysis.

Events are required to contain exactly three lepton candidates satisfying the selection criteria described above. To ensure that the 
trigger efficiency is well determined, at least one of the candidate leptons is required to have $\pt > 25$~\GeV{} and to be geometrically 
matched to a lepton that was selected by the trigger. 

To suppress background processes with at least four prompt leptons, events with a fourth lepton candidate 
satisfying looser selection criteria are rejected. For this looser selection, the $\pt$ of the leptons is 
lowered to $\pt>7$~\GeV{} and ``loose'' identification requirements are used for both the electrons and muons. 
The isolation requirement uses ID track information only and is less stringent.

Candidate events are required to have at least one pair of leptons of the same flavour and of opposite charge, with an invariant mass 
that is consistent with the nominal \Zboson\ boson mass~\cite{Agashe:2014kda} to within $10$~\gev. This pair is considered to be the 
$Z$ boson candidate. If more than one pair can be formed, the pair whose invariant mass is closest to the nominal $Z$ boson mass is 
taken as the \Zboson\ boson candidate. 

The remaining third lepton is assigned to the $W$ boson decay. The transverse mass of the $W$ candidate, computed using \met and the 
\pt of the associated lepton, is required to be greater than $30$~\GeV.

Backgrounds originating from misidentified leptons are suppressed by requiring the lepton associated with the $W$ boson to satisfy more 
stringent selection criteria. Thus, the transverse momentum of these leptons is required to be greater than 20~\GeV. Furthermore, electrons 
associated with the $W$ boson decay are required to pass the ``tight'' likelihood identification requirement~\cite{ATL-PHYS-PUB-2015-041}, 
which has an overall efficiency of  85\%. Finally, these electrons must also pass a tighter isolation requirement, tuned for an efficiency 
of at least 90\%~(99\%) for $\pT>25~(60)$~\GeV.

\section{Background estimation}
\label{sec:BackgroundEstimation}

The background sources are classified into two groups: events where at least one of the candidate leptons is not a prompt 
lepton (reducible background) and events where all candidates are prompt leptons or are produced in the decay of a $\tau$ 
(irreducible background). Candidates that are not prompt leptons are called also ``misidentified'' or ``fake'' leptons.

The reducible background, which represents about half of the total backgrounds, originates from $Z+j$, $Z\gamma$, $t\bar{t}$, 
$Wt$ and $WW$ production processes, with $Z+j$ and $Z\gamma$ being the dominant component (83\%). The reducible backgrounds 
are estimated using data-driven techniques. 
The background from events with two or three fake leptons, {\textit{e.g.}}, from $W+jj$ and multijet processes, is negligible.

Backgrounds from $\ttbar$, $Wt$ and $WW+j$ events (called ``top-like'' in the following) are estimated by exploiting the 
different-flavour decay channels of these processes. These events are categorised based on whether the misidentified lepton 
is an electron or muon. The former are estimated in a control region containing $e^{\pm}\mu^{\mp}e^{\pm}$ events and the 
latter in a $\mu^{\pm}e^{\mp}\mu^{\pm}$ control region. Events in the control regions satisfy the selection criteria described 
in Section~\ref{sec:EventSelection}, except that a different-flavour, opposite-charge lepton pair is associated with the $Z$ boson, 
and the $m_{\ell\ell}$ requirement is removed to increase the number of events. These requirements suppress the contamination 
of events with a leptonically decaying $Z/\gamma^{\ast}$. The dominant contribution is due to top-like processes (75\%). The MC 
predictions for other processes are subtracted from the observed yield. The ratios of observed to expected top-like events in the control regions are  
$0.5\pm0.3$ for misidentified electrons and $1.4\pm0.5$ for misidentified muons, where the uncertainties are due to the statistical 
uncertainties of the data and MC events in the control regions. These ratios are applied to the estimated contributions from 
the simulated top-like backgrounds in the final $W^{\pm}Z$ selection. The kinematic shapes of the top-like background are taken 
from MC simulation for the purposes of control distributions and the exclusive jet multiplicity differential cross-section 
calculation. The shape of the jet multiplicity distribution in the top-like control regions is well modelled by the MC simulation.

Backgrounds from $Z+j$ and $Z\gamma$ processes are estimated by defining a three-lepton $Z$ control sample in data, 
where two of the leptons, referred to as tight (T), meet all identification and isolation criteria described in 
Section~\ref{sec:EventSelection}, and the remaining lepton, referred to as loose (L), fails these requirements and 
instead satisfies less restrictive ones. Events in the $Z$ control sample must satisfy all other \wz\ selection criteria. 
The $Z$ control sample is split into three categories of events, labelled as $N_{\text{LTT}}$, $N_{\text{TLT}}$ and $N_{\text{TTL}}$, 
where the first index refers to the $W$ lepton, and the second and third indexes refer to the higher- and lower-\pt 
leptons from the $Z$ boson decay. The observed number of events in each of these categories is 1535, 61 and 204, 
respectively. The contribution from $Z+j$ and $Z\gamma$ events is greater than 75\%. Processes with at least three 
prompt leptons are subtracted using the MC prediction. This includes the subtraction of $W^{\pm}Z$ events, for which 
the MC prediction is increased by 15\% to agree with previous measurements~\cite{ATLASWZ_8TeV}. The subtraction of 
top-like processes (18\%) uses a procedure with a control region containing one loose lepton, analogous to the procedure described above.

The $Z+j$ and $Z\gamma$ background in the final \wz\ selection is obtained by scaling the observed number of events 
in the $Z$ control sample by an extrapolation factor called the ``fake factor''. The fake factor is measured in a 
data sample with two tight leptons associated with the $Z$ boson and one additional lepton that can be loose or tight. 
To enrich the sample in $Z+j$ and $Z\gamma$ events, the $m_{\text{T}}^{W}$ requirement is reversed and the missing 
transverse momentum is required to be less than 40~\GeV. The fake factor is calculated as the ratio of the number of 
events with a tight third lepton to the number of events with a loose third lepton. The dominant contribution ($>97\%$) 
to the denominator of the fake-factor ratio originates from $Z+j$ and $Z\gamma$ events. Simulation shows that the 
relative fractions of these two processes are similar in this region and the $Z$ control sample where the fake factor 
is applied, justifying the use of a single fake factor to describe both backgrounds. Processes with at least three prompt 
leptons contaminate the events in the numerator of the ratio, particularly at high lepton $\pt$, and are subtracted as 
described for the $Z$ control sample. The fake factor is computed in bins of $\pt$ of the lepton not associated with the 
$Z$ boson, separately for muons and electrons, and considering the different selection criteria used for leptons in the 
analysis. MC simulation is used to verify that the fake factors do not depend on the jet multiplicity of the event. 
The fake-factor values range between $0.02$ and $0.1$.

In brief, the $Z+j$ and $Z\gamma$ estimate in each lepton $\pt$ bin is obtained by extrapolating from events in the $Z$ 
control sample using the following formula:

\begin{equation}
N_{Z+j/Z\gamma} = \left(N_{\text{LTT}} - N_{\text{LTT}}^{\text{prompt}} - N_{\text{LTT}}^{\text{top}}\right) F_{W} + \left(N_{\text{TLT}} -  N_{\text{TLT}}^{\text{prompt}} - N_{\text{TLT}}^{\text{top}}\right)F_{Z} + \left( N_{\text{TTL}} -  N_{\text{TTL}}^{\text{prompt}}  - N_{\text{TTL}}^{\text{top}}\right) F_{Z},
\end{equation}

where $F_{W}$ and $F_{Z}$ denote the fake factors for $W$ and $Z$ leptons, $N_{\text{LTT}}^{\text{prompt}}$, 
$N_{\text{TLT}}^{\text{prompt}}$ and $N_{\text{TTL}}^{\text{prompt}}$ denote the MC prediction of processes with at least 
three prompt leptons, and $N_{\text{LTT}}^{\text{top}}$, $N_{\text{TLT}}^{\text{top}}$ and $N_{\text{TTL}}^{\text{top}}$ 
denote the estimate of top-like events. Both the normalisation and the kinematic shapes of the $Z+j$ and $Z\gamma$ 
background are estimated from the data using this methodology. The estimate of the $Z+j$ and $Z\gamma$ background is 
validated in a subset of the signal region containing events with $30 < m_\textrm{T}^W < 50$ \GeV\ and \met < 40 \GeV, 
which is enriched in background processes.

The reducible background was also assessed with an alternative procedure, the matrix method, used in the previous 
measurement of \wz\ production at $\sqrt{s} = 8$~\TeV{} from the ATLAS Collaboration~\cite{ATLASWZ_8TeV}. The results 
agree with the estimates described above within 5\%. 

Irreducible background events originate from $ZZ$, $t\bar{t} +V $, $VVV$ (where $V$ = $Z$ or $W$), $t Z$ and \wz events 
in which at least one of the bosons decays into leptons via an intermediate $\tau$ decay. The amount of irreducible 
background is estimated using MC simulations. The estimate of the contribution from \wz events decaying via $\tau$-leptons 
is addressed in Section~\ref{sec:CrossSectionDefinition}.

About $70\%$ of the irreducible background is due to $ZZ$ production.  Events from $ZZ$ production survive the  $W^{\pm}Z$ 
event selection either because one lepton falls outside the fiducial volume or because it falls in the fiducial acceptance 
of the detector but is not identified.  The number of $q\bar{q}\to ZZ$ events predicted by \POWHEG is scaled by $1.08$ to 
account for NNLO QCD and NLO EW corrections~\cite{Cascioli:2014yka, Bierweiler:2013dja, Baglio:2013toa}.
The number of $gg\to ZZ$ events predicted by the \SHERPA MC event sample
is scaled by a factor of $1.52$ to account for NLO QCD corrections~\cite{Caola:2015psa}.
These estimates are validated by comparing the MC predictions with the observed event yield, and the distributions of several 
kinematic variables, in a four-lepton data sample enriched in $ZZ$ events. The number of observed events in this validation 
region is 106, with 89\% purity for the $ZZ$ process. Overall agreement between the data and the predictions is within one 
standard deviation of the experimental uncertainty. The shapes of the distributions of the main kinematic variables are also 
found to be well described by the MC predictions. 

\section{Detector-level results}
\label{sec:RecoResults}

Table~\ref{tab:Results:YieldsSummary} summarises the predicted and observed numbers of events together with the
estimated background contributions. The total uncertainties affecting the predicted yields include statistical 
uncertainties,  the theoretical uncertainties in the cross sections as further discussed in Section~\ref{sec:CrossSections}, 
experimental uncertainties discussed in Section~\ref{sec:Systematics} and uncertainty in the integrated luminosity 
for backgrounds estimated using MC predictions. Figure~\ref{fig:Results:WZControlPlots} shows the measured distributions of
the transverse momentum and the invariant mass of the $Z$ candidate, the transverse mass of the $W$ candidate, 
and for the $WZ$ system a variable $m_{\mathrm{T}}^{WZ}$~\cite{ATLASWZ_8TeV} similar to the transverse mass.
The \powhegpythia MC prediction is used for the \wz\ signal contribution. In Figure~\ref{fig:Results:WZControlPlots} 
this contribution is scaled by a global factor of $1.18$ to match the measured inclusive \wz\ cross section in Section~\ref{sec:CrossSections}.
This scaling is only used for an illustrative purpose in this figure and does not affect the measurements.
Figure~\ref{fig:Results:WZControlPlots}  indicates that the MC predictions provide a fair description of the shapes of the data distributions.

\begin{table}[!htbp]
\begin{center}
\begin{tabular}{%
l
S@{$\pm$}S
S@{$\pm$}S
S@{$\pm$}S
S@{$\pm$}S
S@{$\pm$}S
}
\toprule 
{Channel}  &\multicolumn{2}{c}{$eee$} & \multicolumn{2}{c}{$\mu ee$}  & \multicolumn{2}{c}{$e\mu\mu$} & \multicolumn{2}{c}{$\mu\mu\mu$} & \multicolumn{2}{c}{All} \\ 
\midrule 
Data 	& \multicolumn{2}{c}{98}  	& \multicolumn{2}{c}{122} & \multicolumn{2}{c}{166}  	 	& \multicolumn{2}{c}{183}  	& \multicolumn{2}{c}{569}  	\\ 
\midrule 
Total expected 	& 102 & 10  	& 118  &  9 	& 126  &  11	& 160  &  12 	& 506  &  38 	\\ 
\midrule 
$WZ$ 	& 74  &  6 		& 96  &  8 & 97  &  8 	& 129  &  10 	& 396  &  32 	\\ 
$Z+j$, $Z\gamma$ 	& 16  &  7 	 	& 7  &  5 & 14  &  7	& 9  &  5 	& 45  &  17 	\\ 
$ZZ$ 	& 6.7 & 0.7 		& 8.7 & 1.0 	& 8.5 & 0.9 & 11.7 & 1.2 	& 36 & 4 	\\ 
$t\bar{t}+V$ 	& 2.7 & 0.4 		& 3.2 & 0.4 & 2.9 & 0.4 	& 3.4 & 0.5 	& 12.1 & 1.6 	\\ 
$t\bar{t}$, $Wt$, $WW+j$ 	& 1.2 & 0.8 		& 2.0 & 0.9& 2.4 & 0.9  	& 3.6 & 1.5 	& 9.2 & 3.1 	\\ 
$tZ$ 	& 1.28 & 0.20 		& 1.65 & 0.26 & 1.63 & 0.26 	& 2.12 & 0.34 	& 6.7 & 1.1 	\\ 
$VVV$ 	& 0.24 & 0.04 	 	& 0.29 & 0.05 & 0.27 & 0.04	& 0.34 & 0.05 	& 1.14 & 0.18 	\\ 
\bottomrule 
\end{tabular} 
\end{center}
\caption{Observed and expected numbers of events after the \wz\ inclusive selection described in Section~\ref{sec:EventSelection} 
in each of the considered channels and for the sum of all channels. The expected number of \wz\ events from \powhegpythia and the 
estimated number of background events from other processes are detailed. The total uncertainties quoted include the statistical 
uncertainties,  the theoretical uncertainties in the cross sections, the experimental uncertainties and the uncertainty in the 
integrated luminosity.
}
\label{tab:Results:YieldsSummary}
\end{table}

\begin{figure}[!htbp]
\begin{center}
\subfloat[]{\includegraphics[width=.4\textwidth]{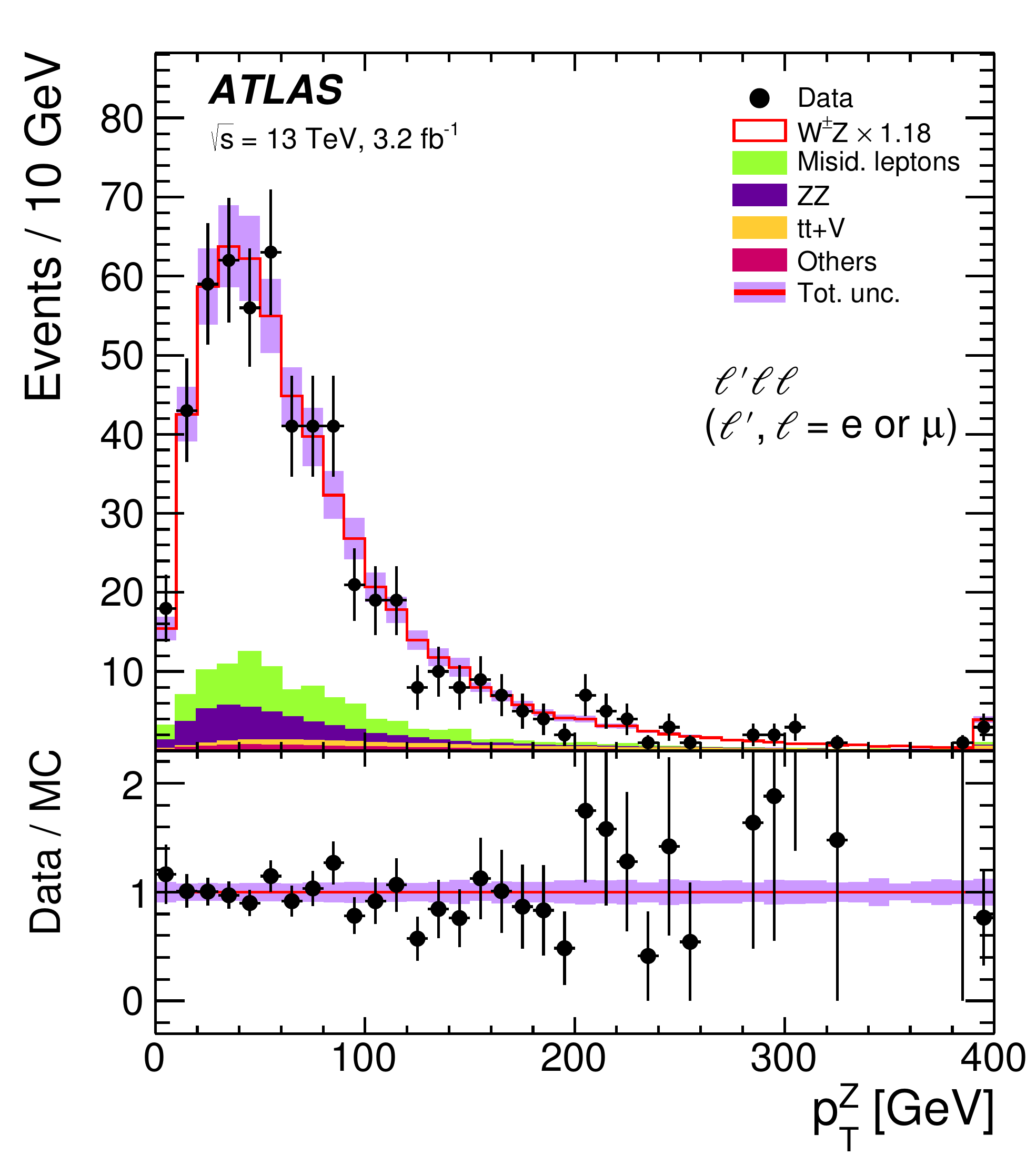}}
\subfloat[]{\includegraphics[width=.4\textwidth]{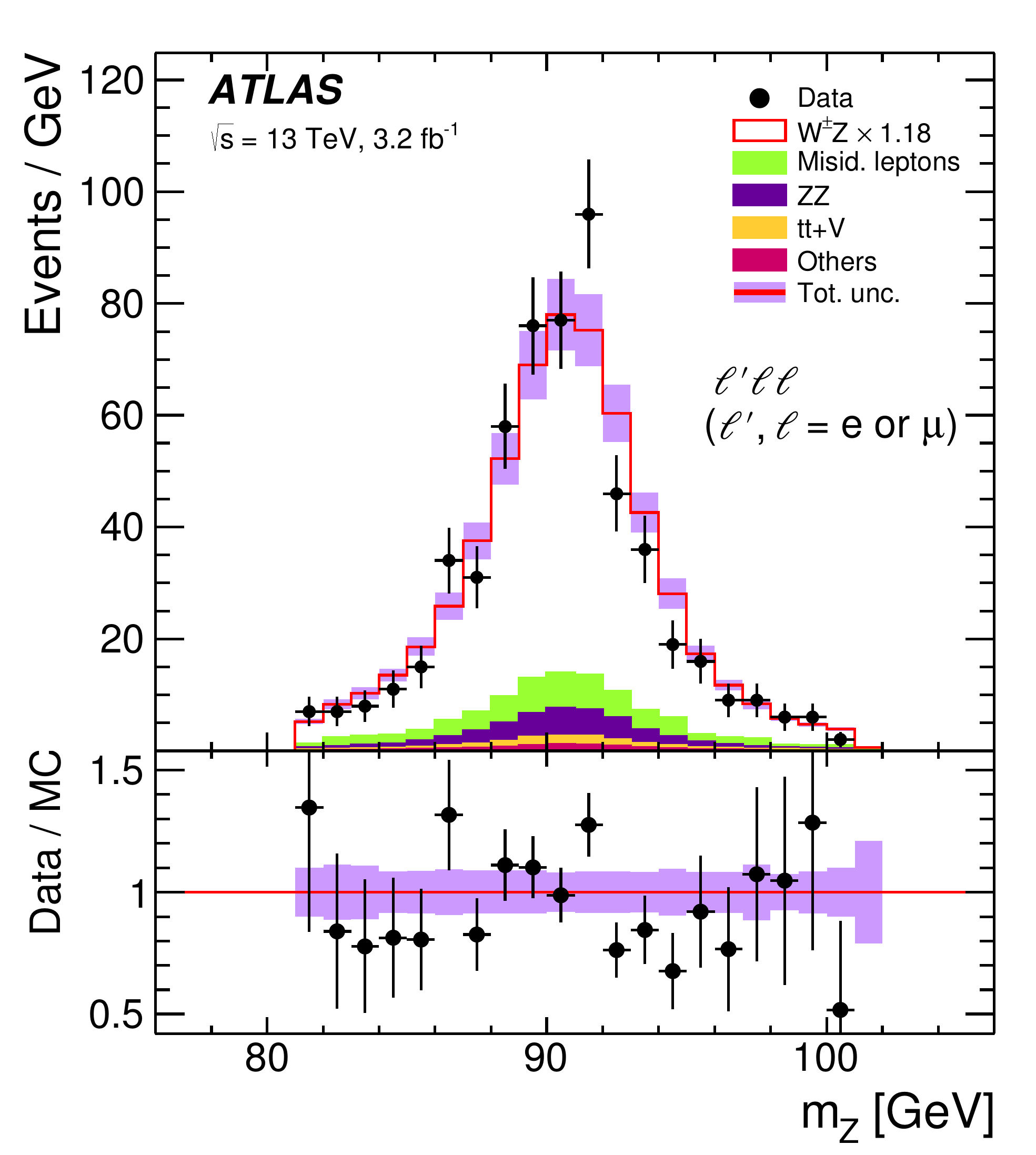}}\\
\subfloat[]{\includegraphics[width=.4\textwidth]{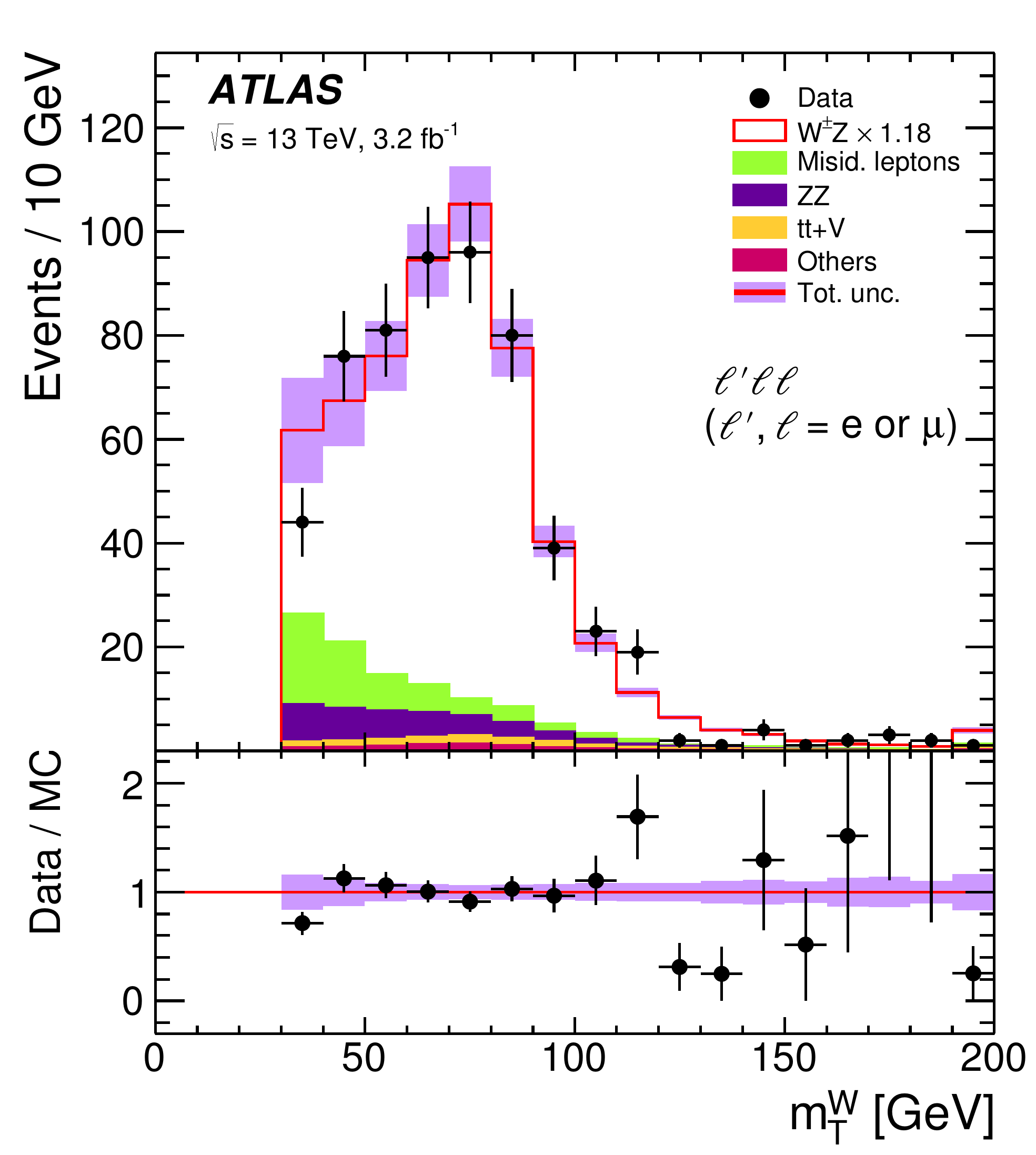}}
\subfloat[]{\includegraphics[width=.4\textwidth]{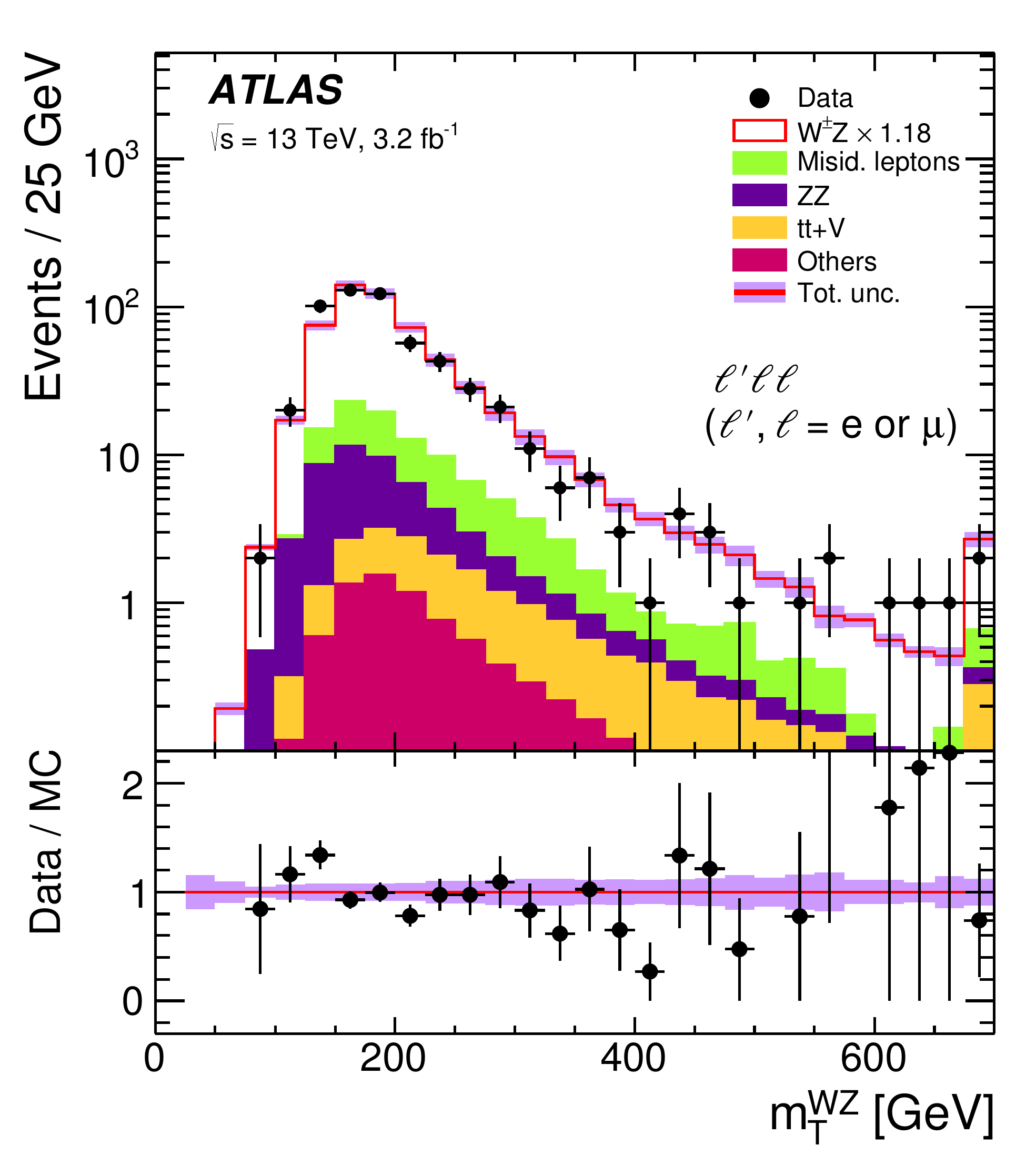}}
\end{center}
 \caption{The distributions for the sum of all channels of the kinematic variables (a) the transverse momentum 
of the reconstructed $Z$ boson $p_\textrm{T}^Z$, (b) the reconstructed $Z$ boson mass $m_Z$, (c) the transverse mass of the
reconstructed $W$ boson $m_\textrm{T}^W$ and (d) the transverse mass variable $m_\textrm{T}^{WZ}$ for the $WZ$ system.
The points correspond to the data, and the histograms correspond to the predictions of the different SM processes. 
All Monte Carlo predictions are scaled to the integrated luminosity of the data using the predicted MC cross sections of each sample.
The sum of the background processes with misidentified leptons is labelled ``Misid. leptons''.
The \powhegpythia MC prediction is used for the \wz\ signal contribution. 
It is scaled by a global factor of $1.18$ to match the measured inclusive \wz\ cross section.
The open red histogram shows the total prediction; the shaded violet band
is the total uncertainty of this prediction. The last bin contains the overflow.
The lower panels in each figure show the ratio of the data points to the
open red histogram with their respective uncertainties.
}
\label{fig:Results:WZControlPlots}
\end{figure}

\section{Corrections for detector effects and acceptance} 
\label{sec:CrossSectionDefinition}

For a given channel \wz\ $\rightarrow \ell^{'\pm} \nu \ell^+ \ell^-$, where $\ell$ and $\ell^{'}$ are either an electron or a
muon,  the integrated fiducial cross section, which includes the leptonic branching fractions of the
$W$ and $Z$, is calculated as

\begin{equation}\label{eq:sigma_fid}
\sigma^{\mathrm{fid.}}_{W^\pm Z \rightarrow \ell^{'} \nu \ell \ell}  = \frac {  N_{\textrm{data}} - N_{\textrm{bkg}} }
 {   \mathcal{L}  \cdot  C_{WZ} }  \times  \left( 1 - \frac {N_\tau} {N_{\textrm{all}} } \right) \, ,
 \end{equation}

where  $N_{\textrm{data}}$ is the number of observed events, $N_{\textrm{bkg}}$ is the estimated number of background events, 
$\mathcal{L}$ is the integrated luminosity and $C_{WZ}$, obtained from simulation, is the ratio of the number of 
selected signal events at detector level to the number of events at particle level in the fiducial phase space defined after QED FSR. 
This factor corrects for detector efficiency and resolution effects and for QED FSR effects. 
The term in parentheses represents the correction applied to the measurement to account for the $\tau$-lepton contribution to the analysis phase space. 
This contribution is estimated using the simulation, from the ratio of  $N_\tau$, the number of selected 
events in which at least one of the bosons decays into a $\tau$ lepton, and $N_{\textrm{all}}$, the number of selected  $WZ$ events with 
decays into any lepton.

The $C_{WZ}$ factors for the $W^- Z$,  $W^+ Z$ and $W^\pm Z$ inclusive processes, as well as the $\tau$-lepton contribution 
to the analysis phase space, $ {N_\tau}/{N_{\textrm{all}}}$, are computed with \powhegpythia for each of the four leptonic 
channels and are shown in Table~\ref{tab:cwz}.
   
\begin{table}[!htbp]
\begin{center}
\begin{tabular}{ccccc}
\toprule
Channel &   $C_{W^-Z}$ &  $C_{W^{+}Z}$  & $C_{W^\pm Z}$ & $ {N_\tau}/{N_{\textrm{all}}}$ \\
\midrule
$eee$           & $0.428$ $\pm$ $0.005$  & $0.417$ $\pm$ $0.004$  & $0.421$ $\pm$ $0.003$ & $0.040$ $\pm$ $0.001$ \\ 
$\mu{ee}$      	& $0.556$ $\pm$ $0.006$  & $0.550$ $\pm$ $0.005$  & $0.553$ $\pm$ $0.004$ & $0.038$ $\pm$ $0.001$ \\ 
${e}\mu\mu$  	& $0.550$ $\pm$ $0.006$  & $0.553$ $\pm$ $0.005$  & $0.552$ $\pm$ $0.004$ & $0.036$ $\pm$ $0.001$ \\ 
$\mu\mu\mu$ 	& $0.729$ $\pm$ $0.007$  & $0.734$ $\pm$ $0.006$  & $0.732$ $\pm$ $0.005$ & $0.040$ $\pm$ $0.001$\\ 
\bottomrule
\end{tabular}
\end{center}
\caption{The $C_{WZ}$ and ${N_\tau}/{N_{\textrm{all}}}$ factors for each of the $eee$, $\mu{ee}$, ${e}\mu\mu$, and $\mu\mu\mu$ inclusive channels. 
The  \powhegpythia MC event sample with the ``resonant shape'' lepton assignment algorithm at particle level
is used. Only statistical uncertainties are reported. }
\label{tab:cwz}
\end{table}

The total cross section is calculated as

\begin{equation}\label{eq:sigma_tot}
\sigma^{\mathrm{tot.}}_{W^\pm Z}  = \frac{  \sigma^{\mathrm{fid.}}_{W^\pm Z \rightarrow \ell^{'} \nu \ell \ell}  } { \mathcal{B}_W \, \mathcal{B}_Z  \, A_{WZ} } \, ,  
\end{equation}

where $ \mathcal{B}_W = 10.86 \pm 0.09 \, \%$ and $\mathcal{B}_Z = 3.3658 \pm 0.0023 \, \%$ are the $W$ and $Z$ leptonic branching fractions~\cite{Agashe:2014kda}, respectively, and
$A_{WZ}$ is the acceptance factor calculated at particle level as the ratio of the number of events in the fiducial
phase space to the number of events in the total phase space as defined in Section~\ref{sec:FiducialPS}.
  
A single acceptance factor of $A_{WZ}$ = $0.343$ $\pm$ $0.002$~(stat.) is obtained using the \powhegpythia simulation by averaging the acceptance factors computed 
in the  $\mu ee$ and $e \mu\mu$ channels.
The use of these channels avoids the ambiguity arising from the assignment at particle level of final-state leptons to the $W$ and $Z$ bosons.
Cross-section differences between $\ell'\ell\ell$ and $\ell\ell\ell$ channels caused by interference effects due to the three identical leptons in the $\ell\ell\ell$ final states are shown by simulation to be below $1\%$.

The differential detector-level distribution of the exclusive jet multiplicity  is corrected for detector resolution and for QED FSR effects using
an iterative Bayesian unfolding method~\cite{D'Agostini:1994zf,RooUnfold}.
Simulated signal events from \powhegpythia are used to obtain a response matrix that accounts 
for bin-to-bin migration effects between the reconstructed and particle-level distribution.

\section{Systematic uncertainties}
\label{sec:Systematics}

The systematic uncertainties in the measured cross sections are due to experimental and theoretical uncertainties 
in the acceptance, uncertainties in the correction procedure for detector effects, uncertainties in the background 
estimation and uncertainties in the luminosity.
  
The theoretical systematic uncertainties in the $A_{WZ}$ and $C_{WZ}$ factors are evaluated by taking 
into account the uncertainties related to the choice of PDF set, to the QCD renormalisation 
$\mu_\text{R}$ and factorisation $\mu_\text{F}$ scales and to the parton showering simulation.
The uncertainties due to the choice of PDF are computed using the CT10 eigenvectors and the envelope
of the differences between the CT10 and CT14~\cite{Dulat:2015mca}, MMHT2014~\cite{Harland-Lang:2014zoa} 
and NNPDF 3.0~\cite{Ball:2014uwa} PDF sets, according to the PDF4LHC recommendations~\cite{PDF4LHC}.
The QCD scale uncertainties are estimated by varying $\mu_\text{R}$ and $\mu_\text{F}$ by factors of 
two around the nominal scale $m_{WZ}/2$ with the constraint $0.5 \leq \mu_\text{R} /\mu_\text{F} \leq 2$, 
where $m_{WZ}$ is the invariant mass of the $WZ$ system.
Uncertainties arising from the choice of parton shower model are obtained from Ref.~\cite{ATLASWZ_8TeV}.              
None of the three sources of theoretical uncertainty have a significant effect on the $C_{WZ}$ factors. 
The uncertainty in the acceptance factor $A_{WZ}$\ is less than $0.5\%$ due to PDF choice, and less than 
$0.7\%$ due to QCD scale choice.
 
The uncertainty in the unfolded jet multiplicity distribution arising from the MC modelling  of the response matrix
in the unfolding procedure is estimated by reweighting the simulated events at particle level to match the unfolded 
results obtained as described in Section~\ref{sec:CrossSectionDefinition}. An alternative response matrix is defined 
using these reweighted MC events and is used to unfold the \powhegpythia reconstructed events.
The systematic uncertainty is estimated by comparing this unfolded distribution to the original particle-level 
\powhegpythia prediction. The size of this uncertainty is at most 15\%. 
            
The experimental systematic uncertainty in the $C_{WZ}$ factors and in the unfolding procedure  includes uncertainties 
in the scale and resolution of the electron energy, muon momentum, jet energy and \met, as well as uncertainties in the 
scale factors applied to the simulation in order to reproduce the   trigger, reconstruction, identification and isolation 
efficiencies measured in data. The uncertainties in the jet energy scale are obtained from $\sqrt{s} = 13$~\TeV{} 
simulations and {\textit{in situ}} measurements, similar to the ones described in Ref.~\cite{ATL-PHYS-PUB-2015-015}. 
The uncertainty in the jet energy resolution is derived by extrapolating measurements in Run-1 data to $\sqrt{s} = 13$~\TeV.
The uncertainty in the \MET is estimated by propagating the uncertainties in the transverse momenta of hard physics objects and
by applying momentum scale and resolution uncertainties to the track-based soft term. The uncertainty associated with pile-up 
modelling is of the order of $1\%$ and can reach up to $2.9\%$ in the 0-jet bin of the unfolded jet multiplicity distribution.
For the measurements of the $W$ charge-dependent cross sections, an uncertainty arising from the charge misidentification 
of leptons is also considered. It affects only electrons and leads to an uncertainty of less than $0.05\%$ in the ratio of 
$W^+Z$ to $W^- Z$ integrated cross sections determined by combining the four decay channels.

The dominant contribution among the experimental systematic uncertainties in the $eee$ and $\mu ee$  channels is due to 
the uncertainty in the electron identification efficiency, contributing at most 1.4\% uncertainty to the integrated cross 
section, while in the $e \mu\mu$ and $\mu\mu\mu$ channels it originates from the muon reconstruction efficiency and is at 
most  $1.1\%$. The systematic uncertainties in the measured cross sections are determined by repeating the analysis after 
applying appropriate variations for each source of systematic uncertainty to the simulated samples.

The dominant uncertainty in the reducible $t\bar{t}$, $Wt$ and $WW+j$ background arises from the number of data and MC events 
in the control regions used to estimate the top-like processes and amounts to 32\% of the estimated yield. An uncertainty of 
$2\%$ is assigned due to the extrapolation from the control regions to the \wz\ signal region.

The statistical precision of $3\%$ in the reducible $Z+j$ and $Z\gamma$ background estimate is determined by the size of the 
$N_{\text{LTT}}$, $N_{\text{TLT}}$ and $N_{\text{TTL}}$ categories in the $Z$ control sample. Uncertainties due to the size 
of the sample used to derive the fake factor amount to $21\%$ of the estimated $Z+j$ and $Z\gamma$ yield. An uncertainty of 
$15\%$ is assigned to the contributions from processes with at least three prompt leptons, which are subtracted from the
sample used to derive the fake factor. This has an $18\%$ impact on the $Z+j$ and $Z\gamma$ estimate. The uncertainty due to 
the subtraction of $\ttbar$, $Wt$ and $WW$ processes is smaller than $2\%$. To account for differences between the region in 
which the fake factor is calculated and the $Z$ control sample where it is applied, including the different relative contributions 
from $Z+j$ and $Z\gamma$ processes in each region, the fake factor is calculated using MC events in both regions, and the full 
difference between the two is taken as a systematic uncertainty, representing 26\% of the estimated $Z+j$ and $Z\gamma$ yield.
Overall, the $Z+j$ and $Z\gamma$ background is estimated with a precision of $38\%$.

A theoretical uncertainty in the $ZZ$ cross section of $8\%$~\cite{Cascioli:2014yka, Bierweiler:2013dja, Baglio:2013toa,Caola:2015psa} 
is assigned as a global uncertainty in the amount of $ZZ$ background predicted by the MC simulation. An additional uncertainty 
of $3\%$ to $6\%$ is assigned due to the correction applied to $ZZ$ MC events with unidentified leptons. 

The uncertainty due to other irreducible background sources is evaluated by propagating the uncertainty in their MC cross sections. 
These are $13\%$~($12\%$) for $\ttbar W$~($\ttbar Z$)~\cite{Alwall:2014hca}, $20\%$ for $VVV$~\cite{ATL-PHYS-PUB-2016-002} and $15\%$ 
for $tZ$~\cite{ATLASWZ_8TeV}.
 
An uncertainty in the integrated luminosity of $2.1\%$ is applied to the signal normalisation as well as to all background contributions 
that are estimated purely using MC simulations. The uncertainty is derived following a methodology similar to that detailed in 
Refs.~\cite{LumiUncertainty,LumiUncertainty2}, from a calibration of the luminosity scale using $x$-$y$ beam-separation scans performed 
in August $2015$. It has an effect of $2.4\%$ on the measured cross sections.
 
The total systematic uncertainty in the \wz fiducial cross section, excluding the luminosity uncertainty, varies between 
$4\%$ and $10\%$ for the four different measurement channels, and is dominated by the uncertainty in the reducible background estimate. 
The statistical uncertainty in the fiducial cross-section measurement is slightly larger than the systematic uncertainty.
Table~\ref{tab:SystematicSummary} shows the statistical uncertainty and main sources of systematic 
uncertainty in the \wz fiducial cross section for each of the four channels and their combination.
  
\begin{table}[!htbp]
\begin{center}
\begin{tabular}{l S S S S S} 
\toprule 
 & {$e e e$} & {$\mu e e$}  & {$e \mu\mu$} & {$\mu\mu\mu$} & {Combined} \\ 
\midrule 
\multicolumn{6}{c}{Relative uncertainties [\%]}\\ 
\midrule 
$e$ energy scale & 0.5 & 0.2 & 0.3 & <0.1 & 0.2 \\ 
$e$ id. efficiency & 1.4 & 1.1 & 0.6 & {---} & 0.7 \\ 
$\mu$ momentum scale & < 0.1 & < 0.1 & < 0.1 & 0.1 & < 0.1 \\ 
$\mu$ id. efficiency & {---} & 0.6 & 1.0 & 1.4 & 0.7 \\ 
$E_{\mathrm{T}}^{\mathrm{miss}}$ and jets & 0.3 & 0.4 & 0.8 & 0.7 & 0.6 \\ 
Trigger & < 0.1 & 0.1 & 0.1 & 0.2 & 0.1 \\ 
Pile-up & 0.7 & 1.1 & 1.0 & 0.7 & 0.9 \\ 
Misid. lepton background & 10 & 4.6 & 4.8 & 3.2 & 3.6 \\ 
$ZZ$ background & 1.0 & 0.7 & 0.6 & 0.7 & 0.7 \\ 
Other backgrounds & 0.5 & 0.5 & 0.3 & 0.3 & 0.4 \\ 
\midrule 
Uncorrelated& 2.2 & 1.3 & 1.4 & 1.7 & 0.8 \\ 
\midrule 
Total sys. uncertainty & 11 & 5.1 & 5.3 & 4.1 & 4.1 \\ 
Luminosity& 2.4 & 2.4 & 2.3 & 2.3 & 2.4 \\ 
Statistics& 14 & 11 & 10 & 8.8 & 5.1 \\ 
\midrule 
Total& 18 & 12 & 11 & 10 & 7.0 \\ 
\bottomrule
\end{tabular} 
\end{center}
\caption{Summary of the relative uncertainties in the measured fiducial cross section $\sigma^{\textrm{fid.}}_{W^\pm Z}$ for each channel 
and for their combination. The uncertainties are reported as percentages. The decomposition of the total systematic uncertainty into the 
main sources correlated between channels and the source uncorrelated between channels is indicated in the first rows.
}
\label{tab:SystematicSummary}
\end{table}

\section{Cross-section measurements}
\label{sec:CrossSections}

The measured fiducial  cross sections in the four channels are combined using a $\chi^2$ minimisation method that accounts for correlations 
between the sources of systematic uncertainty affecting each channel~\cite{Glazov:2005rn,Aaron:2009bp,Aaron:2009wt}. The combination of the 
\wz cross sections in the fiducial phase space yields a total $\chi^2$ per degree of freedom ($n_{{\rm dof}}$) of $\chi^2/n_{{\rm dof}} = 6.9/3$.
The combinations of the $W^+ Z$ and the $W^- Z$ cross sections separately yield $\chi^2/n_{{\rm dof}} = 5.3/3$ and $2.0/3$, respectively.
 
Combining the four channels to obtain a weighted mean value, the cross section of \wz production and decay to a single leptonic channel with 
muons or electrons in the detector fiducial region is

\begin{equation}
\sigma_{W^\pm Z \rightarrow \ell^{'} \nu \ell \ell}^{\mathrm{fid.}} =  63.2~\pm~ 3.2 \, \textrm{(stat.)}~\pm~2.6 \, \textrm{(sys.)}~\pm 1.5 \, \textrm{(lumi.) fb}.
\end{equation}   

The SM NLO QCD prediction from \powhegpythia is $53.4  ^{+1.6}_{-1.2} \,\textrm{(PDF)} ^{+2.1}_{-1.6} \,\textrm{(scale)} \; \rm{fb}$.
The theoretical predictions are estimated using the CT10 PDF set and setting the dynamic QCD scales, $\mu_\text{R}$ and $\mu_\text{F}$, 
equal to $m_{WZ}/2$. The uncertainty in the theoretical prediction due to the PDF is estimated using the eigenvectors of the CT10 PDF 
set scaled to $68$\% confidence level (CL) and the envelope of the differences between the results obtained with the CT14~\cite{Dulat:2015mca}, 
MMHT2014~\cite{Harland-Lang:2014zoa} and NNPDF3.0~\cite{Ball:2014uwa} NLO PDF sets. The QCD scale uncertainty is estimated conventionally 
by varying the scales $\mu_{\mathrm{R}}$ and $\mu_{\mathrm{F}}$ by factors of two around the nominal value of $m_{WZ}/2$ with the constraint 
$0.5 \leq \mu_{\mathrm{R}} /\mu_{\mathrm{F}} \leq 2$. The measured \wz production cross sections are compared to the SM NLO prediction 
from \powhegpythia in Figure~\ref{fig:xsectionperchannel} and all results for $W^\pm Z$, $W^+Z$ and $W^-Z$ final states are reported in 
Table~\ref{tab:XSection:FiducialCSAll}. The measured cross section is larger than the SM prediction, as also were the corresponding 
cross-section measurements performed at lower centre-of-mass energies by the ATLAS Collaboration~\cite{Aad:2012twa,ATLASWZ_8TeV}.

Using the integrated fiducial cross section measured for $W^\pm Z$ production at $\sqrt{s} = 8$~\TeV{} from Ref.~\cite{ATLASWZ_8TeV}, 
the ratio $\sigma_{W^\pm Z}^{\textrm{fid.}, 13 \, \textrm{\TeV}} / \sigma_{W^\pm Z}^{\textrm{fid.}, 8 \, \textrm{\TeV}}$ of the $W^\pm Z$ 
production cross sections at the two centre-of-mass energies of $8$ and $13$~\TeV{} is calculated and yields

\begin{equation}
\frac{\sigma_{W^\pm Z}^{\textrm{fid.}, 13 \, \textrm{\TeV}} }{ \sigma_{W^\pm Z}^{\textrm{fid.}, 8 \, \textrm{\TeV}} }  =  1.80 \pm 0.10 \; \textrm{(stat.)} \pm 0.08 \; \textrm{(sys.)} \pm 0.06 \; \textrm{(lumi.)}. 
\end{equation}

All uncertainties are treated as uncorrelated between the measurements at the two beam energies.
The measured ratio is in good agreement with the Standard Model prediction of $1.78 \pm 0.03$ from \powhegpythia.

\begin{table}[!htbp]
\begin{center}
\begin{tabular}{lccccc}
\toprule
Channel   &  $\sigma^{\mathrm{fid.}}$ & $\delta_{\mathrm{stat.}}$ & $\delta_{\mathrm{sys.}}$ & $\delta_{\mathrm{lumi.}}$ &$\delta_{\mathrm{tot.}}$\\
          &  [fb] & [\%] &  [\%] & [\%] &[\%]\\
\midrule
\multicolumn{6}{c}{~}\\ [-12.0pt]
\multicolumn{6}{c}{$\sigma^{\mathrm{fid.}}_{W^{\pm} Z \rightarrow \ell^{'} \nu \ell \ell}$}\\ [4.0pt]
\midrule
$e^\pm{ee}$                                         &  50.5    &       14.2 &  10.6 &           2.4 &  17.8\\   
$\mu^\pm{ee}$                                      &  55.1    &      11.1 &  5.1 &           2.4 & 12.4\\ 
$e^\pm\mu\mu $                                 &  75.2   &       9.5 &  5.3 &            2.3 &  11.1\\ 
$\mu^\pm\mu\mu$                              &  63.6   &         8.9 &  4.1 &              2.3 &  10.0\\ 
\midrule
Combined      & 63.2    &          5.2 &  4.1 &              2.4 &  7.0\\ 
\midrule 
SM prediction  & 53.4    &          --- &  --- &            --- &  6.0\\ 
\midrule 
\multicolumn{6}{c}{~}\\ [-12.0pt]
\multicolumn{6}{c}{$\sigma^{\mathrm{fid.}}_{W^{+} Z \rightarrow \ell^{'} \nu \ell \ell}$}\\ [4.0pt]
\midrule
$e^+ee$                                            & 28.0    &          19.2 &            11.2 &             2.4 &  22.3\\   
$\mu^+{ee}$                                        & 32.2    &          14.4 &             5.0 &              2.4 & 15.3\\ 
$e^+\mu\mu$                                   & 45.0    &          12.1 &             4.6 &               2.3 & 13.1\\ 
$\mu^+\mu\mu$                               & 36.5    &           11.6 &             4.1 &               2.3 &  12.5\\ 
\midrule
Combined          & 36.7    &           6.7 &             3.9 &                2.3 & 8.1\\ 
\midrule 
SM prediction  & 31.8    &          --- &  --- &            --- &  5.8\\ 
\midrule 
\multicolumn{6}{c}{~}\\ [-12.0pt]
\multicolumn{6}{c}{$\sigma^{\mathrm{fid.}}_{W^{-} Z \rightarrow \ell^{'} \nu \ell \ell}$}\\ [4.0pt]
\midrule
$e^-ee$                                            & 22.5    &           21.0 &            10.5 &                2.4 &     23.6\\   
$\mu^-{ee}$                                        & 22.9   &            17.5 &            5.8 &                2.4 &     18.5 \\ 
$e^-\mu\mu$                                   & 30.2   &            15.2 &            6.9 &                2.3 &    16.8\\ 
$\mu^-\mu\mu$                                & 27.1   &            13.7 &            5.0 &               2.4&         14.7 \\  
\midrule
Combined           & 26.1   &             8.1 &            4.7 &                2.4 &        9.6\\ 
\midrule
SM prediction  & 21.6    &          --- &  --- &            --- &  7.9\\ 
\bottomrule 
\end{tabular}
\caption{Fiducial integrated cross section in fb, for $W^\pm Z$, $W^+ Z$ and $W^- Z$ production, measured in each of the $eee$, $\mu{ee}$, ${e}\mu\mu$, and $\mu\mu\mu$ channels and all four channels combined.
The statistical ($\delta_{\mathrm{stat.}}$), total systematic ($\delta_{\mathrm{sys.}}$), luminosity ($\delta_{\mathrm{lumi.}}$) and total ($\delta_{\mathrm{tot.}}$) uncertainties are given in percent.
}
\label{tab:XSection:FiducialCSAll}
\end{center}
\end{table}

The ratio of $ W^+Z$ to $W^-Z$ production cross sections is

\begin{equation}
\frac{\sigma_{W^{+}Z \rightarrow \ell^{'} \nu \ell \ell}^{\textrm{fid.}}}{\sigma_{W^{-}Z \rightarrow \ell^{'} \nu \ell \ell}^{\textrm{fid.}}}  =  1.39\pm 0.14 \,\textrm{(stat.)} \pm 0.03 \,\textrm{(sys.)}.
\end{equation}

Most of the systematic uncertainties, and especially the luminosity uncertainty, cancel in the ratio, and the measurement is dominated by the statistical uncertainty.
The measured cross-section ratios, for each channel and for their combination, are compared in Figure~\ref{fig:WPMXSection:WpWmRatio} 
to the SM prediction of $1.47 ^{+0.03}_{-0.06}$, which is calculated with \powhegpythia and the CT10 PDF set.

\begin{figure}[!htbp]
\begin{center}
\includegraphics[width=11cm]{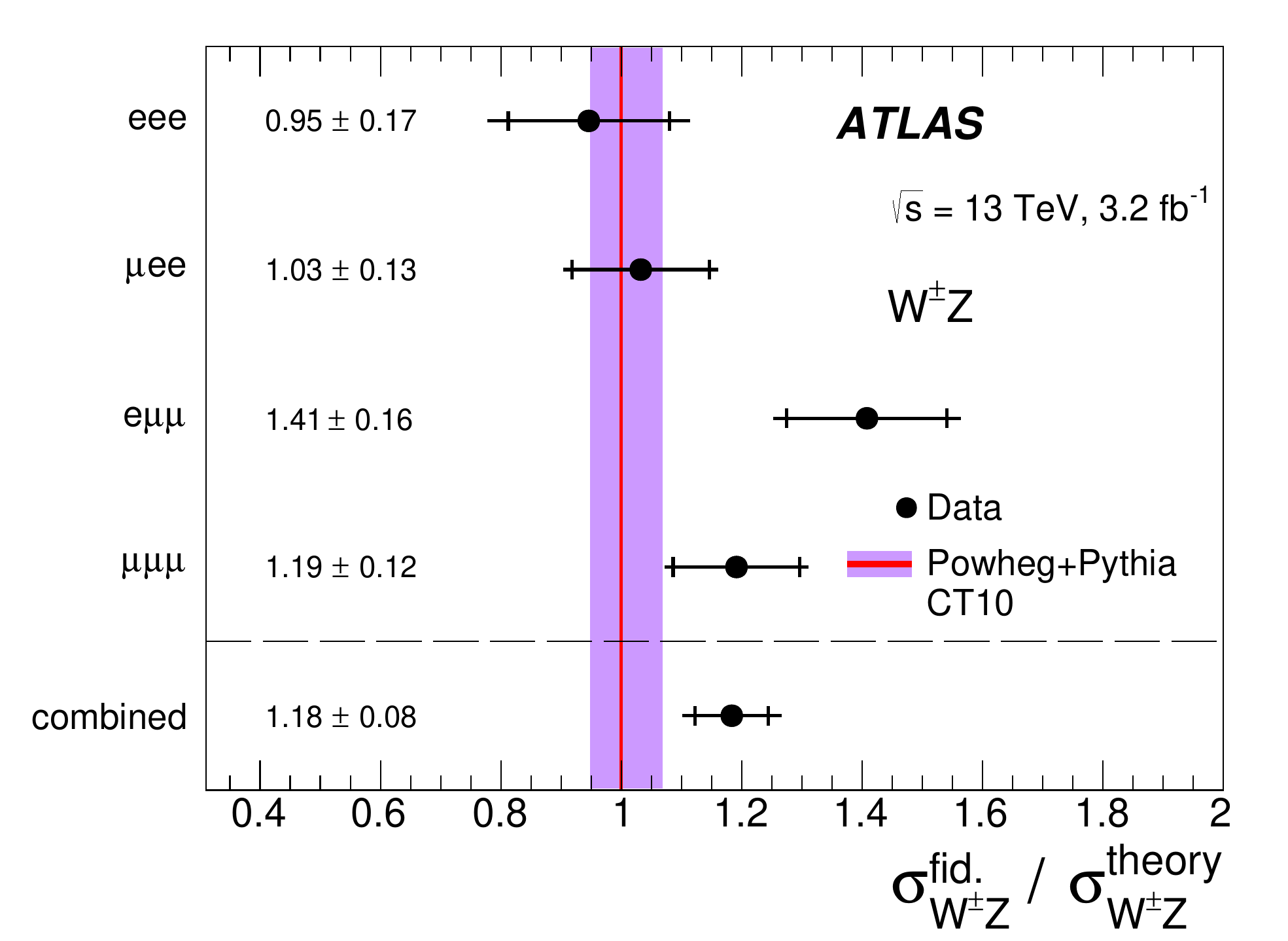}
\caption{Ratio of the measured \wz\ integrated cross sections in the fiducial phase space to the NLO SM prediction from 
\powhegpythia in each of the four channels and for their combination. The inner and outer error bars on the data points 
represent the statistical and total uncertainties, respectively. The NLO SM prediction from \powhegpythia using the CT10 
PDF set is represented by the red line; the shaded violet band is the total uncertainty in this prediction.
}
\label{fig:xsectionperchannel}
\end{center}
\end{figure}

\begin{figure}[!htbp]
\begin{center}
\includegraphics[width=11cm]{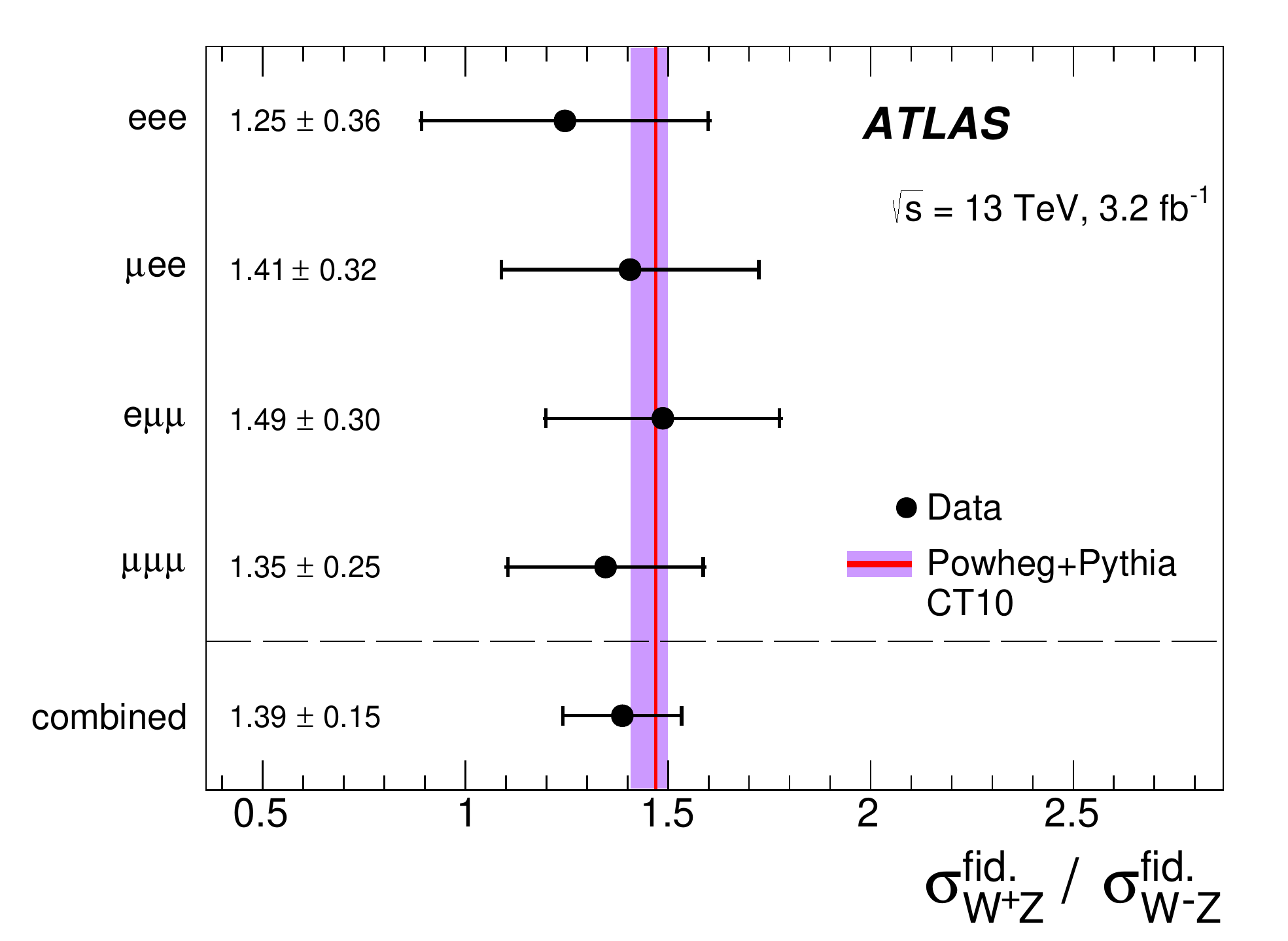}     
\caption{Measured ratios $\sigma^{\mathrm{fid.}}_{W^{+}Z} / \sigma^{\mathrm{fid.}}_{W^{-}Z}$ of $W^{+}Z$ and $W^{-}Z$ 
integrated cross sections in the fiducial phase space in each of the four channels and for their combination.
The error bars on the data points represent the total uncertainties, which are dominated by the statistical uncertainties.
The NLO SM prediction from \powhegpythia using the CT10 PDF set is represented by the red line; the shaded violet band
is the total uncertainty in this prediction.
}

\label{fig:WPMXSection:WpWmRatio}
\end{center}
\end{figure}

The combined fiducial cross section is extrapolated to the total phase space. The result is

\begin{equation}
\sigma_{W^{\pm}Z}^{\rm{tot.}}  =  50.6 \pm 2.6 \,\rm{(stat.)} \pm 2.0 \,\rm{(sys.)} \pm 0.9 \,\rm{(th.)} \pm 1.2 \,\rm{(lumi.) \, pb},
\end{equation}

where the theoretical uncertainty accounts for the uncertainties in the $A_{WZ}$ factor due to the choice of PDF set, QCD scales and parton shower model.
The NLO SM prediction calculated with \powhegpythia is $42.4  \pm 0.8 \,\rm{(PDF)} \pm 1.6 \,\rm{(scale)} \; \rm{pb}$.
A recent calculation~\cite{Grazzini:2016swo} of the \wz production cross section at NNLO in QCD with MATRIX, obtained using the NNPDF3.0 PDF set 
and with $\mu_\text{R}$ and $\mu_\text{F}$ scales fixed to $(m_W+m_Z)/2$, yields $48.2^{+1.1}_{-1.0} \,\textrm{(scale)} \; \rm{pb}$, which is in better agreement with the measurement.
As this prediction does not include effects of QED final-state radiation, a correction factor of 0.972 as estimated from \powhegpythia is applied.

Finally, the exclusive jet multiplicity cross section is presented in Figure~\ref{fig:unfoldResultNjetMjj} and compared to the 
predictions from \powhegpythia and \sherpa. The shape of the measured cross section as a function of jet multiplicity is 
described well by \SHERPA, but it is reproduced poorly by \powhegpythia. The matrix-element calculation in the \SHERPA 
prediction includes up to three jets at LO, while in the \powhegpythia prediction only the leading jet is included, 
and higher jet multiplicities are described by the parton shower models.

\begin{figure}[!htbp]
\begin{center}
\includegraphics[width=0.5\textwidth]{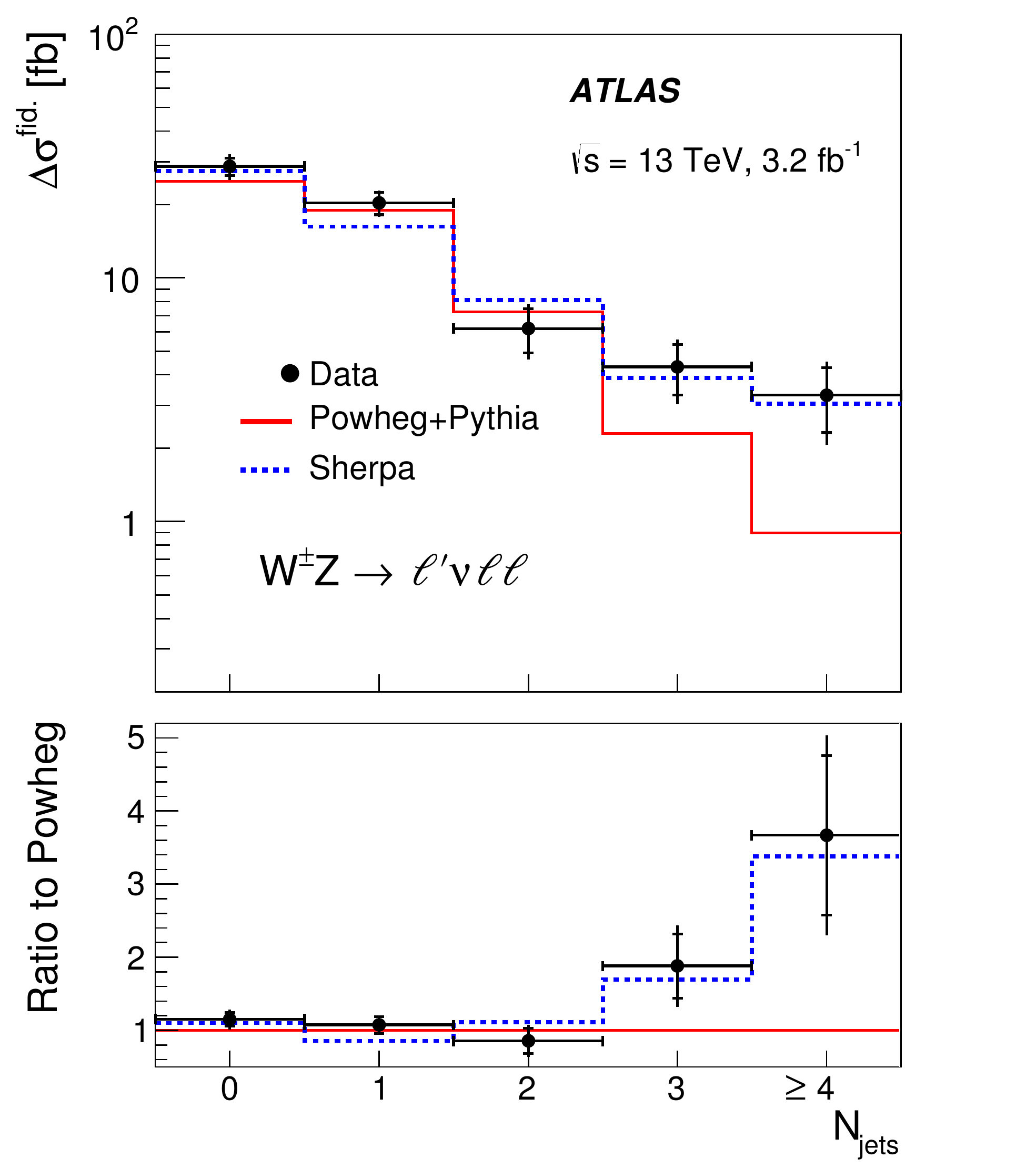}
\caption{The measured $W^{\pm}Z$ differential cross section in the fiducial phase space as a function of the exclusive jet 
multiplicity of jets with $p_\textrm{T} > 25$~\GeV. The inner and outer error bars on the data points represent the statistical 
and total uncertainties, respectively. The measurements are compared to the prediction from \powhegpythia (red line) and \sherpa (dashed blue line).
}
\label{fig:unfoldResultNjetMjj}
\end{center}
\end{figure}

\section{Conclusion}
\label{sec:conclusion}

Measurements of \wz\ production in $\sqrt{s} = 13$~\TeV{} $pp$ collisions at the LHC are presented.
The data were collected with the ATLAS detector in $2015$ and correspond to an integrated luminosity of $3.2$~fb$^{-1}$.
The measurements use leptonic decay modes of the gauge bosons to electrons or muons and are performed 
in a fiducial phase space closely matching the detector acceptance.
The measured inclusive cross section in the fiducial region for one leptonic decay channel is 
$\sigma_{W^\pm Z \rightarrow \ell^{'} \nu \ell \ell}^{\mathrm{fid.}} =  63.2~\pm~ 3.2 \, \textrm{(stat.)}~\pm~2.6 \, \textrm{(sys.)}~\pm 1.5 \, \textrm{(lumi.) fb}$.
The NLO Standard Model prediction from \powhegpythia is $53.4^{+3.6}_{-2.8}$~fb. 
The measured cross section is higher than the SM NLO prediction; a similar excess was found in the 
cross-section measurements performed at lower centre-of-mass energies by the ATLAS Collaboration. 

The ratio of the measured cross sections at the two centre-of-mass energies yields
$\sigma_{W^\pm Z}^{\textrm{fid.}, 13 \, \textrm{\TeV}} / \sigma_{W^\pm Z}^{\textrm{fid.}, 8 \, \textrm{\TeV}} =  1.80 \pm 0.10 \; \textrm{(stat.)} \pm 0.08 \; \textrm{(sys.)} \pm 0.06 \; \textrm{(lumi.)}$, 
in good agreement with the SM NLO prediction of $1.78 \pm 0.03$ from \powhegpythia.
The $W^+Z$ and $W^-Z$ production cross sections are measured separately in the fiducial phase space and are 
reported; their ratio is 
$\sigma_{W^{+}Z \rightarrow \ell^{'} \nu \ell \ell}^{\textrm{fid.}}/\sigma_{W^{-}Z \rightarrow \ell^{'} \nu \ell \ell}^{\textrm{fid.}} = 1.39\pm 0.14 \,\textrm{(stat.)} \pm 0.03 \,\textrm{(sys.)}$.
This result is in agreement with the SM NLO expectation from \powhegpythia of $1.47^{+0.03}_{-0.06}$. 
The measured cross section extrapolated to the total phase space is 
$50.6 \pm 2.6 \,\rm{(stat.)} \pm 2.0 \,\rm{(sys.)} \pm 0.9 \,\rm{(th.)} \pm 1.2 \,\rm{(lumi.) \, pb},$
in very good agreement with the SM NNLO prediction from MATRIX of $48.2^{+1.1}_{-1.0} \,\textrm{(scale)} \; \rm{pb}$.

Finally, the \wz production cross section is measured as a function of the exclusive jet multiplicity
and compared to the SM predictions of \powhegpythia and \sherpa. The \sherpa prediction is found to provide 
a better description of the data, at low and high jet multiplicities.

\section*{Acknowledgements}
\input{acknowledgements/Acknowledgements}



\printbibliography


\newpage \input{atlas_authlist}

%
%

\end{document}

%% file: mydocument-metadata.tex
\AtlasTitle{Measurement of the \wz boson pair-production cross section in $pp$ collisions at $\rts = 13\tev$ with the ATLAS Detector}

\author{The ATLAS Collaboration}
\PreprintIdNumber{CERN-PH-2016-108}
\AtlasJournalRef{Phys. Lett. B 762 (2016) 1}
\AtlasDOI{10.1016/j.physletb.2016.08.052}

\AtlasAbstract{%
The production of \wz events in proton--proton collisions at a centre-of-mass energy of $13$~\TeV\ 
is measured with the ATLAS detector at the LHC. The collected data correspond to an integrated 
luminosity of {\mbox {$3.2$~fb$^{-1}$}}.
The \wz\ candidates are reconstructed using leptonic decays of the gauge bosons into electrons or muons.
The measured inclusive cross section in the detector fiducial region for leptonic decay modes is 
$\sigma_{W^\pm Z \rightarrow \ell^{'} \nu \ell \ell}^{\textrm{fid.}} =  63.2~\pm~ 3.2 \, \textrm{(stat.)}~\pm$\linebreak$~2.6 \, \textrm{(sys.)}\pm 1.5 \, \textrm{(lumi.)}$~fb.
In comparison, the next-to-leading-order Standard Model prediction is $53.4^{+3.6}_{-2.8}$~fb. 
The extrapolation of the measurement from the fiducial to the total phase space yields 
$\sigma_{W^{\pm}Z}^{\rm{tot.}}  =  50.6 \pm 2.6 \,\rm{(stat.)} \pm 2.0 \,\rm{(sys.)} \pm 0.9 \,\rm{(th.)} \pm 1.2 \,\rm{(lumi.)}$~pb, 
in agreement with a recent next-to-next-to-leading-order calculation of $48.2^{+1.1}_{-1.0}$~pb. 
The cross section as a function of jet multiplicity is also measured, together with the charge-dependent $W^+Z$ and $W^-Z$ cross sections and their ratio. 
}

%% file: acknowledgements/Acknowledgements.tex

We thank CERN for the very successful operation of the LHC, as well as the
support staff from our institutions without whom ATLAS could not be
operated efficiently.

We acknowledge the support of ANPCyT, Argentina; YerPhI, Armenia; ARC, Australia; BMWFW and FWF, Austria; ANAS, Azerbaijan; SSTC, Belarus; CNPq and FAPESP, Brazil; NSERC, NRC and CFI, Canada; CERN; CONICYT, Chile; CAS, MOST and NSFC, China; COLCIENCIAS, Colombia; MSMT CR, MPO CR and VSC CR, Czech Republic; DNRF and DNSRC, Denmark; IN2P3-CNRS, CEA-DSM/IRFU, France; GNSF, Georgia; BMBF, HGF, and MPG, Germany; GSRT, Greece; RGC, Hong Kong SAR, China; ISF, I-CORE and Benoziyo Center, Israel; INFN, Italy; MEXT and JSPS, Japan; CNRST, Morocco; FOM and NWO, Netherlands; RCN, Norway; MNiSW and NCN, Poland; FCT, Portugal; MNE/IFA, Romania; MES of Russia and NRC KI, Russian Federation; JINR; MESTD, Serbia; MSSR, Slovakia; ARRS and MIZ\v{S}, Slovenia; DST/NRF, South Africa; MINECO, Spain; SRC and Wallenberg Foundation, Sweden; SERI, SNSF and Cantons of Bern and Geneva, Switzerland; MOST, Taiwan; TAEK, Turkey; STFC, United Kingdom; DOE and NSF, United States of America. In addition, individual groups and members have received support from BCKDF, the Canada Council, CANARIE, CRC, Compute Canada, FQRNT, and the Ontario Innovation Trust, Canada; EPLANET, ERC, FP7, Horizon 2020 and Marie Sk{\l}odowska-Curie Actions, European Union; Investissements d'Avenir Labex and Idex, ANR, R{\'e}gion Auvergne and Fondation Partager le Savoir, France; DFG and AvH Foundation, Germany; Herakleitos, Thales and Aristeia programmes co-financed by EU-ESF and the Greek NSRF; BSF, GIF and Minerva, Israel; BRF, Norway; Generalitat de Catalunya, Generalitat Valenciana, Spain; the Royal Society and Leverhulme Trust, United Kingdom.

The crucial computing support from all WLCG partners is acknowledged gratefully, in particular from CERN, the ATLAS Tier-1 facilities at TRIUMF (Canada), NDGF (Denmark, Norway, Sweden), CC-IN2P3 (France), KIT/GridKA (Germany), INFN-CNAF (Italy), NL-T1 (Netherlands), PIC (Spain), ASGC (Taiwan), RAL (UK) and BNL (USA), the Tier-2 facilities worldwide and large non-WLCG resource providers. Major contributors of computing resources are listed in Ref.~\cite{ATL-GEN-PUB-2016-002}.

%% file: atlas_authlist.tex
\begin{flushleft}
{\Large The ATLAS Collaboration}

\bigskip

M.~Aaboud$^\textrm{\scriptsize 135d}$,
G.~Aad$^\textrm{\scriptsize 86}$,
B.~Abbott$^\textrm{\scriptsize 113}$,
J.~Abdallah$^\textrm{\scriptsize 64}$,
O.~Abdinov$^\textrm{\scriptsize 12}$,
B.~Abeloos$^\textrm{\scriptsize 117}$,
R.~Aben$^\textrm{\scriptsize 107}$,
O.S.~AbouZeid$^\textrm{\scriptsize 137}$,
N.L.~Abraham$^\textrm{\scriptsize 149}$,
H.~Abramowicz$^\textrm{\scriptsize 153}$
% ...
\bigskip
\\
% ...
$^{12}$ Institute of Physics, Azerbaijan Academy of Sciences, Baku, Azerbaijan\\
% ...
$^{64}$ University of Iowa, Iowa City IA, United States of America\\
% ...
$^{86}$ CPPM, Aix-Marseille Universit{\'e} and CNRS/IN2P3, Marseille, France\\
% ...
$^{107}$ Nikhef National Institute for Subatomic Physics and University of Amsterdam, Amsterdam, Netherlands\\
% ...
$^{113}$ Homer L. Dodge Department of Physics and Astronomy, University of Oklahoma, Norman OK, United States of America\\
% ...
$^{117}$ LAL, Univ. Paris-Sud, CNRS/IN2P3, Universit{\'e} Paris-Saclay, Orsay, France\\
% ...
$^{135}$ $^{(a)}$ Facult{\'e} des Sciences Ain Chock, R{\'e}seau Universitaire de Physique des Hautes Energies - Universit{\'e} Hassan II, Casablanca; $^{(b)}$ Centre National de l'Energie des Sciences Techniques Nucleaires, Rabat; $^{(c)}$ Facult{\'e} des Sciences Semlalia, Universit{\'e} Cadi Ayyad, LPHEA-Marrakech; $^{(d)}$ Facult{\'e} des Sciences, Universit{\'e} Mohamed Premier and LPTPM, Oujda; $^{(e)}$ Facult{\'e} des sciences, Universit{\'e} Mohammed V, Rabat, Morocco\\
% ...
$^{137}$ Santa Cruz Institute for Particle Physics, University of California Santa Cruz, Santa Cruz CA, United States of America\\
% ...
$^{149}$ Department of Physics and Astronomy, University of Sussex, Brighton, United Kingdom\\
% ...
$^{153}$ Raymond and Beverly Sackler School of Physics and Astronomy, Tel Aviv University, Tel Aviv, Israel\\
% ...
$^{a}$ Also at Department of Physics, King's College London, London, United Kingdom\\
$^{b}$ Also at Institute of Physics, Azerbaijan Academy of Sciences, Baku, Azerbaijan\\
$^{c}$ Also at Novosibirsk State University, Novosibirsk, Russia\\
$^{d}$ Also at TRIUMF, Vancouver BC, Canada\\
$^{e}$ Also at Department of Physics {\&} Astronomy, University of Louisville, Louisville, KY, United States of America\\
% ...
\end{flushleft}
